\documentclass{article}

\usepackage{arxiv}
\usepackage[utf8]{inputenc} 
\usepackage[T1]{fontenc}    
\usepackage{hyperref}       
\usepackage{url}            
\usepackage{booktabs}       
\usepackage{amsfonts}       
\usepackage{nicefrac}       
\usepackage{microtype}      
\usepackage{lipsum}
\usepackage{graphicx}

\usepackage{tabulary}
\usepackage{booktabs}
\usepackage{todonotes}

\usepackage[utf8]{inputenc}
\usepackage{algorithm}
\usepackage[noend]{algpseudocode}

\graphicspath{ {./images/} }

\title{Integration of Unpaired Single-cell Chromatin Accessibility and Gene Expression Data via Adversarial Learning}

\author{
 Yang Xu \\
  UT-ORNL Graduate School of Genome Science and Technology\\
  The University of Tennessee\\
  \texttt{yxu71@vols.utk.edu} \\
   \And
 Andrew Jeremiah Strick \\
  The University of Tennessee\\
  \texttt{astrick@vols.utk.edu} \\
}

\begin{document}
\maketitle

\begin{abstract}
Deep learning has empowered analysis for single-cell sequencing data in many ways and has generated deep understanding about a range of complex cellular systems. As the booming single-cell sequencing technologies brings the surge of high dimensional data that come from different sources and represent cellular systems with different features, there is an equivalent rise and challenge of integrating single-cell sequence across modalities. Here, we present a novel adversarial approach to integrate single-cell chromatin accessibility and gene expression data in a semi-supervised manner. We demonstrate that our method substantially improves data integration from a simple adversarial domain adaption approach, and it also outperforms two state-of-the-art (SOTA) methods.
\end{abstract}

\section{Introduction}
Within the last decade, single-cell sequencing technologies have advanced our understanding on a broad range of biological systems\cite{gawad2016single}. These technologies revealed distinct cellular compositions, and analyses based on these single-cell sequencing data also provided reliable database for biomedical research and valuable reference for medical discovery. Coming with these technology breakthroughs, Deep-learning-based single-cell analysis also gained great attention in recent years and has been used to address a range of difficulties, including accurate cell-type annotation\cite{ma2020actinn}, gene expression imputation\cite{arisdakessian2019deepimpute}, and doublet identification\cite{bernstein2020solo}. In these tasks, deep learning showed striking advantages. For example, the automatic and accurate annotation via a deep classification model saves researchers from manual cell-type annotation\cite{ma2020actinn},\cite{lopez2018deep},\cite{kimmel2020scnym}. Another application of deep learning in single-cell gene expression data is imputation and denoising of gene expression. Though there has been a dramatic improvement in single-cell RNA-seq technology, the problem of zero-inflation still remains as a grand challenge\cite{lahnemann2020eleven}. Due to the difficulty of modeling technical zero values and biological zeros, deep learning became a more appealing alternative for this task. An autoencoder (AE) is a common artificial neural network that learns lower dimension representation for high dimensional data in an unsupervised manner. Both DeepImpute\cite{arisdakessian2019deepimpute} and DCA\cite{eraslan2019single} adopted a variant of AE model to impute and denoise single-cell gene expression data. These approaches and many others are revealing the power of deep learning applied to single-cell sequencing data. 

Among these applications, data integration is a rising challenge in single-cell analysis, as increasing numbers of single-cell sequencing data become available, and the types of sequencing data become more diverse\cite{stuart2019integrative}. However, there are few comprehensive tools to address the challenge of data integration across modalities. In integration of single-cell gene expression data, batch effects are usually a prominent variation when these data come from multiple sources. Batch effects can be due to multiple factors, for example experiments conducted by different personnel and data generated by different sequencing machines. Most batch effects are not biologically relevant, and single-cell database confounded by batch effects are also not applicable for general use. Therefore, removing batch effects is a critical step for revealing true biological variation and necessary for building batch-invariant and applicable database. Two research groups almost simultaneously developed the same deep learning approach to address batch effects within single-cell gene expression data\cite{bahrami2020deep},\cite{dincer2020adversarial}. Both methods adopted the generative adversarial network as the main framework for learning latent space that is not entangled with batch effects. The success of integrating multi-source single-cell gene expression data relies in the adversarial domain adaption that assists the model to approximate the joint distribution for data from different sources. Besides integrating single-cell gene expression data, integration of other modalities of single-cell sequencing data is becoming even more important as technology breakthroughs make it possible to capture multiple data types from the same single cell. For example, sci-CAR and SHARE-seq can simultaneously profile chromatin accessibility and gene expression for thousands of single cells\cite{cao2018joint},\cite{ma2020chromatin}. scMethyl-HiC and sn-m3C-seq can profile DNA methylation and 3D chromatin structure at the same time and at single-cell resolution\cite{li2019joint},\cite{lee2019simultaneous}. However, these data types do not naturally share the same feature space: expression data are described using genes as features, while chromatin accessibility data is reported across all genomic locus. Therefore, data integration across modalities is more challenging than integrating multi-source gene expression data. Feature engineering based on biological understanding can solve the issue of feature discrepancy and make single-cell sequencing data have the shared feature space (See Data preparation). However, it won't necessarily bridge the gap between two different modalities. To address data integration across modalities, we are motivated by these two batch-removal studies and transfer adversarial domain adaption to modality alignment\cite{bahrami2020deep},\cite{dincer2020adversarial}. Different from these two previous studies, we further introduce a cylce-consistency network approach to learn the joint representation for both chromatin accessibility and gene expression data\cite{zhu2017unpaired}. In this study, we demonstrate that our method substantially improves accuracy of integration from a simple adversarial domain adaption approach and our novel adversarial approach also outperforms two SOTA methods.

\section{Background}

Deep generative models have been tested in single-cell analysis extensively and demonstrated their efficacy of learning discriminative representation from the original high dimensional space. The most common generative models are Variational Autoencoder (VAE). Variants of VAE models, which differ in their sampling approaches, have been proposed to learn representations for single-cell gene expression data\cite{lopez2018deep},\cite{bahrami2020deep},\cite{wang2019bermuda},\cite{lotfollahi2018generative}. The core component of VAE is the use of reconstruction loss, which encodes a sample in the representation where it is drawn from a certain distribution, for example, a Gaussian distribution. The use of reconstruction loss also has an advantage of mapping noisy data to high-quality data, which further extends the ability of generative model to denoise data or impute gene expression. Starting from a basic deep generative model (VAE), both Bahrami et. al and Dincer et. al introduced adversarial domain loss into the generative model and transferred the learning approach from reconstruction of data to diminishing non-biological variation\cite{bahrami2020deep},{dincer2020adversarial}. This approach turned out to be effective in removing batch effects within single-cell gene expression data.

However, approaches mentioned above are also limited to address batch effects within single-cell gene expression data. Generally, these methods can perform data integration of the same modality, by approximating the joint distribution of single-cell gene expression data from different sources, give that these datasets measure the same biological system. If the two modalities measure the same variation of one biological system but from different views, We reason that a similar adversarial domain adaption approach could be devised to address bimodal data integration. For example, chromatin accessibility (ATAC-seq) and gene expression (RNA-seq) data. Numerous studies have shown that most biological cellular systems represent similar variation in both chromatin accessibility and gene expression views. Simply put, if a biological system can be defined by the variation of gene expression that represents its distinct cellular composition, the variation of chromatin accessibility can also define this biological system with a similar cellular composition. Thus, adversarial domain adaption is aimed to diminish difference between modalities, instead of dealing with batch effects. However, there are only a few generative adversarial network (GAN) models proposed for this task. In this study, we start from the adversarial domain adaption model proposed by  Bahrami et. al and Dincer et. al and further introduce a novel use of cylce-consistency network\cite{bahrami2020deep},\cite{dincer2020adversarial},\cite{zhu2017unpaired}. Our method removes modality differences while keeping true biological variation within the two modalities. Our method also extends use of the adversarial learning, from single modal data integration to bimodal data integration. We expect this work will fill the missing gap of use of GANs in integration of multi-modal single-cell data.

Before we move to details of our model architecture and training, we would like to review a few key studies that our method is built on.

\subsection{Unsupervised representation learning}
Deep metric learning has shown effective representation learning without supervision. In a recent study, Chen et al. used a simple framework to learn visual representations in a self-supervised manner\cite{chen2020simple}. They duplicated each image into two counterparts through image perturbation. The goal of learning is to maximize the consistency of any paired replicates in the latent space $z$. To achieve this goal, they applied the loss function as shown in Eq. 1. In a N-sample batch, there will be $2N$ samples through data augmentation. In fact, each augmented image has a corresponding counterpart which is basically the same, despite of the added image perturbation. Then, $cos$ quantifies the cosine similarity of image $i$ and $j$/$k$ in the latent space $z$. In their study, this simple framework turns out to be a very effective way to learn the discriminative representation without supervision. In our study, we adopted the same approach to learn representation for chromatin accessibility data.

\begin{equation}
l_{i,j} = -log(\frac{exp(cos(z_i,z_j))/\tau}{\sum_{k=1}^{2N}exp(cos(z_i,z_k))/\tau})
\end{equation}

\subsection{Adversarial domain adaption}
Generative models with adversarial domain adaption were successfully shown to transfer targets to source style and has its application in image translation\cite{tzeng2017adversarial}. Recently, both Bahrami et. al and Dincer et. al incorporated adversarial domain adaption into a generative model for removing batch effects within single-cell expression data\cite{bahrami2020deep},\cite{dincer2020adversarial}. For both studies, the goal is to find batch-invariant representation for single-cell gene expression from different sources. To achieve this, they stacked a discriminator to the encoder and trained the discriminator to distinguish which source the cell comes from using the latent space $z$ projected by encoder. Adversarial training, in this case, will push the encoder to approximate the joint distribution and become capable of projecting cells to the integrated representation no matter of where they come from. 

\subsection{Cylce-consistency adversarial network}
Besides the use of adversarial domain adaption in style transferring and image translation, a method called cycleGAN presented a SOTA outcome to transfer image style to another domain, without known its pair during training\cite{zhu2017unpaired}. The success of establishing a connection between two image domains relies in the concept they called "cycle consistency". Starting from the original image, a generator network translates the image to the other domain. Then, a second generator network translates the image back to its original domain. Through this cycle, the translated-back image should be the same as the original image. Based on this information, adversarial training of generators can establish a reversible connection between two image domains. Different from goal of cycleGAN, we aim to learn integrated representation instead of translating chromatin accessibility to gene expression or vice verse. However, the fundamental concept is the same: we establish a cycle from encoder to generator, and from generator back to encoder, and then the cycle-consistency loss is applied at the level of latent space $z$.

\subsection{Interpolation Consistency Training}
In our study, besides learning an integrated representation for both chromatin accessibility and gene expression data, we also aim to transfer labels of gene expression data to predict cell-types for unknown chromatin accessibility data. However, the classifier network is primarily supervised by gene expression data and will be blind to the intrinsic structure of chromatin accessibility. In a study on interpolation consistency training (ICT), Verma et al demonstrated how learning the decision boundary is biased by limited labeled data when the labeled data is not good representative for the whole population\cite{verma2019interpolation}. However, without labels, we can neither use unlabeled data for supervised training. To solve this problem and make full use of available unlabeled data, Verma et al. proposed the ICT and reframed learning approach into a semi-supervised manner. In our study, we reason that gene expression data cannot completely represent chromatin accessibility data. To avoid that the learned decision boundary is biased towards labeled gene expression data, we generated mixed  $\lambda$ * chromatin accessibility + (1- $\lambda$) * gene expression data, and we add an ICT term as regularization to the total loss (as Eq. 2). A simple rational behind can be explained in a way that mixed output from model should be a combination of outputs of the two independent modalities.

\begin{equation}
C(E( \lambda * R + (1- \lambda) * A)) = C'(E'( \lambda * R)) + C'(E'((1 - \lambda) * A))
\end{equation}

\begin{figure*}[t!]
    \includegraphics[width=\textwidth,page=1]{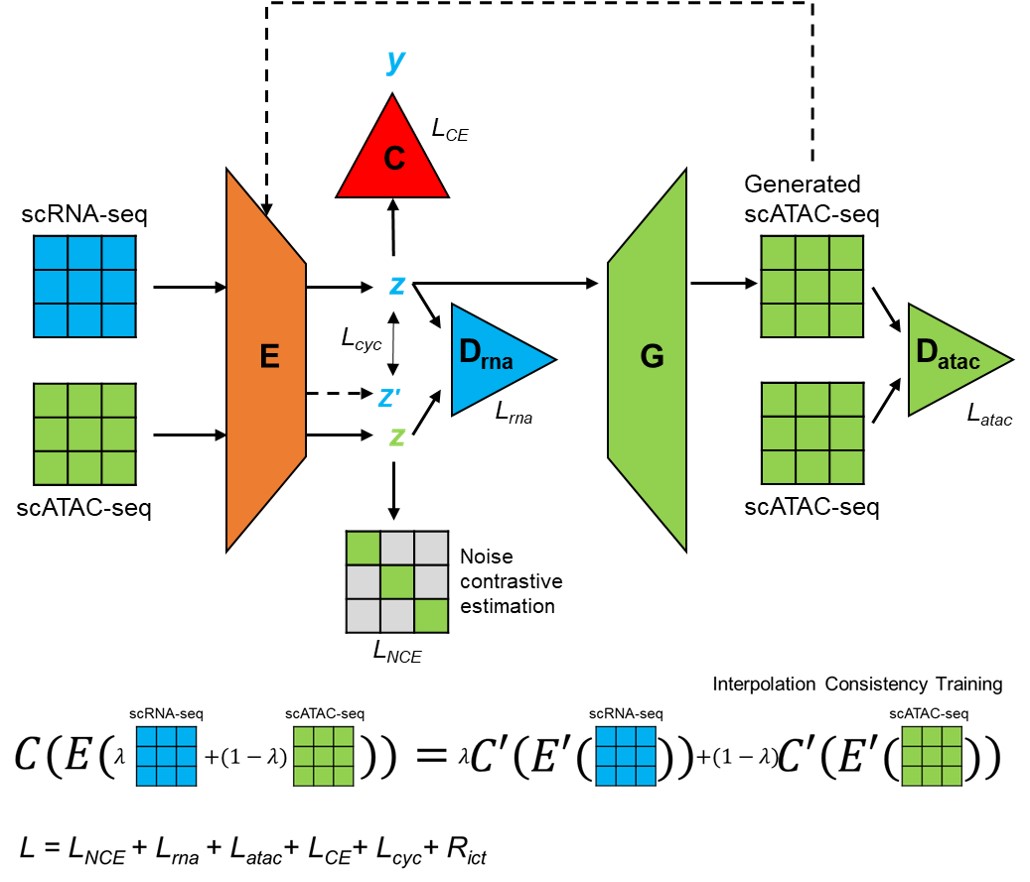}
    \centering
    \caption{Final architecture of adversarial model for integrating single-cell chromatin accessibility and gene expression data. E: encoder network; C: classifier network; Drna and Datac: discriminator network; G: generator network. The final loss term consists of 5 independent loss functions and 1 regularization term through interpolation consistency training.}
\end{figure*}

\section{Methods}
\subsection{Model architecture}
Our final model architecture is shown in Figure 1. Encoder ($E$) serves as a feature extractor that projects both high dimensional chromatin accessibility and gene expression data into the joint low dimension space. Our method consists of representation learning and modality alignment. For representation learning, we take advantage of available annotation of single-cell gene expression data and perform supervised representation learning by training a classifier network ($C$). For chromatin accessibility data, labels are unknown during training. However, to make full use of the chromatin accessibility data, we select an unsupervised metric approach to learn discriminative representation for chromatin accessibility data\cite{chen2020simple}. Meanwhile, we also introduce interpolation consistency training to frame the representation learning as semi-supervised\cite{verma2019interpolation}. For modality alignment, we use two separate discriminator network for two distinct uses. The first discriminator network ($D_{rna}$) is attached to $E$ and is trained with adversarial domain adaption loss. This discriminator aims to distinguish which source the latent space z extracted by $E$ comes from, while $E$ is pushed to learn the joint distribution so that $D_{rna}$ is less able to distinguish the modality source of latent space $z$. The second discriminator network ($D_{atac}$) follows a generator network ($G$) that generates chromatin accessibility data based latent space $z'$ from the gene expression data. Adversarial training here will push the G to find a connection between chromatin accessibility and gene expression data. Since the generated chromatin accessibility data is based on the latent space $z$ of real gene expression data, the new latent space $z'$ of generated data should align with its corresponding $z$ of real gene expression data. Therefore, we can add cylce-consistency loss as demonstrated in cycleGAN method\cite{zhu2017unpaired}.

\begin{algorithm}
	\caption{Training procedure} 
	\begin{algorithmic}[1]
		\For {each epoch}
			\For {$Batch=1,2,\ldots,N$}\Comment{representation learning}
				\State Calculate $L_{ce}$, $NCE$, and $R_{ict}$ 
				\State Update weights in $E$ and $C$
			\EndFor
			\For {$Batch=1,2,\ldots,N$}\Comment{modality alignment}
				\State Calculate $L_{rna}$, $L_{atac}$, $L_{cyc}$
				\State Update weights in $E$, $D_{rna}$, $D_{atac}$ and $G$
			\EndFor
		\EndFor
	\end{algorithmic} 
\end{algorithm}

\subsection{Model detail}
\subsubsection{Discriminative representation learning}
The main component of our model is a single encoder that projects cells from the original gene-based feature space to the joint lower dimension representation. The E consists of two fully-connected layers that have 1024 and 128 nodes, respectively. Each fully-connected layer is followed by a batch normalization layer and both are activated by a ReLu function. To learn a discriminative representation, we take advantage of labels for single-cell gene expression data, since most RNA-seq data are well annotated and it is easy to find annotated single-cell gene expression data. We stack a C to E, which only has one fully-connected layer, and we train cells from single-cell gene expression data in supervised manner. For cells from single-cell chromatin accessibility data, cell-type annotation remains unknown, we choose an unsupervised approach to learn the representation. In this approach, we duplicate each cell into two replicates by adding Gaussian noise. Though different noise are added to the two replicas, they are basically the same cell. There, we apply noise contrastive estimation to guide E to reduce difference between any paired replicates. To avoid that C is biased by gene expression data, we further introduce interpolation consistency training (ICT). The ICT will further take advantage of availability of unlabeled chromatin accessibility data and assist the encoder and the classifier to find optimal decision boundaries for both chromatin accessibility and gene expression data. We sample $\lambda$  with  $\beta$ (0.2, 0.2) distribution and mix $\lambda$ * gene expression data with (1- $\lambda$) * chromatin accessibility data.

\subsubsection{Adversarial learning}
Discriminative representation learning will only approximate the structure of chromatin accessibility and gene expression data independently. However, it nevertheless learns the joint distribution for both modalities. Adversarial learning, in this part, is aimed to align both modalities. We stack the $D_{rna}$ onto the $E$. $D_{rna}$ takes latent space $z$ and produces a single sigmoid activated output. The aim of $D_{rna}$ is to discriminate which modality the cell comes from using latent space $z$ extracted by $E$. Given gene expression modality as label 1 and chromatin accessibility modality as label 0, adversarial domain loss would guide $D_{rna}$ to correctly assign 0 to cells in chromatin accessibility data and 1 to cells in gene expression data, while $E$ tries to diminish the modality difference so as that $D_{rna}$ assigns 1 to all cells, no matter which modality they are from. For adversarial training of $G$ and $D_{atac}$, cells in chromatin accessibility modality would be given label 1 while generated chromatin accessibility data given label 0. $G$ also takes latent space $z$ as input but produces linear activated output with the same shape as chromatin accessibility data. $D_{atac}$ takes the original chromatin accessibility data as input and produces a single sigmoid activated output. The generated chromatin accessibility data is based on latent space $z$ of real gene expression data. Therefore, if we forward generated chromatin accessibility data through $E$ again, the new latent space $z'$ should align with latent space $z$ of the real gene expression data. Based on this information, we use mean squared error to calculate cycle-consistency loss. For all adversarial learning, we simultaneously update weights in $E$ and $D_{rna}$, or $G$ and $D_{atac}$ through gradient reverse\cite{ganin2015unsupervised}.
\section{Experiments}

\subsection{Experimental setup}
\subsubsection{Data selection and preparation}
We selected one unpaired (human PBMC) and one paired (mouse skin) chromatin accessibility/gene expression data. In mouse skin data, though each cell in gene expression modality has its corresponding counterpart in chromatin accessibility, we keep this information as unknown. Therefore, both datasets are unpaired in this study. These two unpaired data measure the same biological systems with chromatin accessibility views and gene expression views, respectively. Cells in both modalities were also manually annotated and validated by authors. Therefore, cell-type annotation of gene expression data can serve as label for supervised learning while cell-type annotation of chromatin accessibility data as ground truth for evaluation. Keep in mind that annotation of chromatin accessibility data is only used for evaluation but not used for training. Data summary is provided in Table 1. Human PBMC data is a relatively small set and only contains 13 cell types. Because of the small size, we chose human PBMC data for method development and examined contribution of each loss function to data integration. Mouse skin data consists 64,462 cells in total that belong to 22 different cell-types. Thus, we selected mouse skin data as the dataset for final examination by comparing our method with two SOTA methods.

\begin{center}
\scalebox{0.8}{
\begin{tabular}{ |c|c|c|c| }
\hline
\multicolumn{4}{|c|}{Table 1. Data summary} \\
\hline
Data & number of cells in ATAC-seq & number of cells in RNA-seq & number of cell-types \\
\hline
Human Peripheral Blood Mononuclear Cell (PBMC) & 9,432 & 7,866 & 13 \\ 
\hline
Mouse skin & 32,231 & 32,231 & 22 \\ 
\hline
\end{tabular}}
\end{center}

To engineer and match features in both chromatin accessibility and gene expression data, we used a common practice that transforms ATAC-seq peak matrix to gene activity matrix. Here, we briefly explain the rationale behind. RNA-seq commonly measures gene expression. A common matrix of single-cell gene expression data would be where each row is one cell and each column is expression values of one gene. The whole matrix represents gene expressions of the whole genome across all cells. Because of this, it is also easy to annotate gene expression data by associating the strong expression signals of marker genes with a cell type. ATAC-seq, on the other hand, quantifies how accessible genomic locus are to certain regulatory molecules. Therefore, a common matrix single-cell chromatin accessibility data would be where each row is one cell (the same as single-cell gene expression data) and each column is accessibility values of one genomic locus. These genomic locus can embed in one gene body, and the sum of accessibility values of all genomic locus within that gene body can represent the potential of transcription. Therefore, conversion involves the summation of all accessibility peaks within each gene body and uses the sum of peaks to represent gene accessibility. The converted gene accessibility matrix would be where each row is one cell, and each column is accessibility values of one gene. After conversion, we can do a simple filtering and reordering to have features of chromatin accessibility and gene expression data matched. This process also avoids using separate encoders to handle inputs from chromatin accessibility and gene expression data.

\subsubsection{Evaluation}
To evaluate integration of chromatin accessibility and gene expression data, we select silhouette coefficient as the metric. We further reduce the learned representation to 2d UMAP space, and we calculate silhouette coefficient based on true cell-type annotation. Silhouette coefficient closer to 1 indicates better preservation of data structure and good mixture of modality.

Since we did not use labels for chromatin accessibility data during training, we can test how well the model predicts cell-type label for chromatin accessibility data. Once the encoder is trained, we first project both modalities into the same latent space through the encoder. Then, we use gene expression data with cell-type labels to train a Support Vector Machine (SVM) classifier. Finally, we use the SVM classifier to predict cell-type label for chromatin accessibility data. Macro and weighted F1 scores are calculated by comparing true cell-type label with the SVM classifier prediction. Higher F1 scores indicates good label transfer from gene expression to chromatin accessibility.

\subsection{Results}

\begin{figure*}[t!]
    \includegraphics[width=\textwidth,height=5cm,page=2]{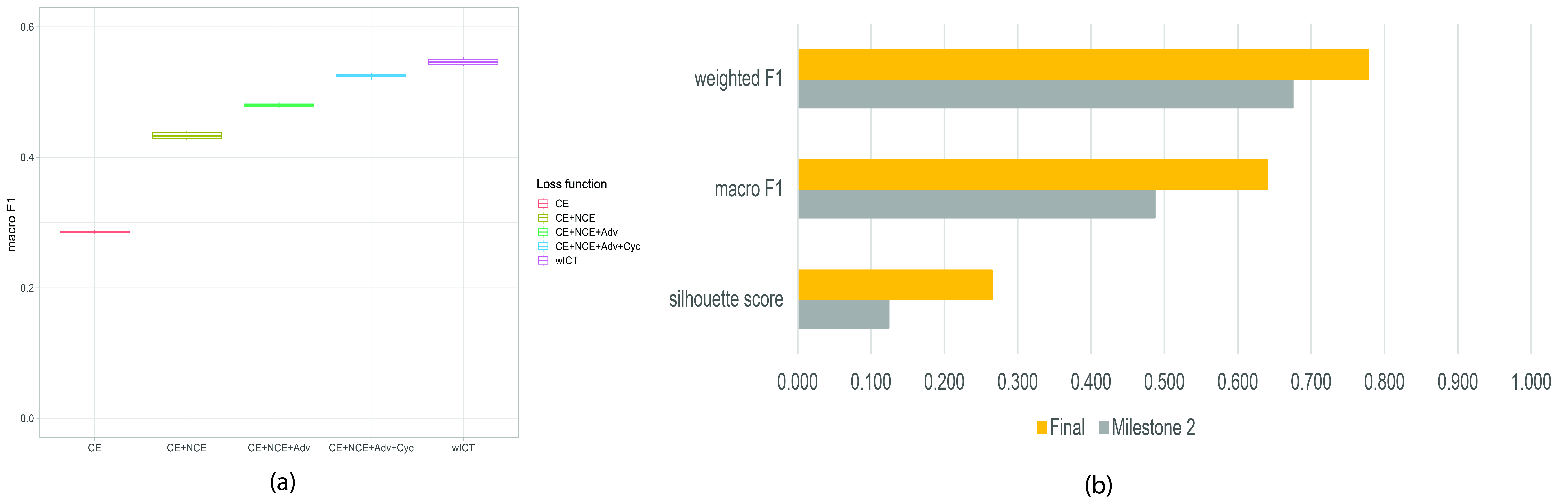}
    \centering
    \caption{(a) Evaluation of loss functions that were used for model training. CE: cross entropy; NCE: noise contrastive estimation; Adv: adversarial domain adaption loss; Cyc: cylce-consistency loss; wICT: with interpolation consistent training. (b) Model improvement. Our final model is compared with our milestone 2 model. Evaluations are based on silhouette score, macro F1 and weighted F1. For all three metrics, higher indicates better performance.}
\end{figure*}

\subsubsection{Establishment of final model}

We first examined how much each loss function used in our final model contribute to accuracy of integration, quantified by macro F1 on human PBMC chromatin accessibility data, shown in Figure 2(a). Our baseline performance is training the model only with cross entropy. In this case, the model uses single-cell gene expression data for supervised learning and doesn't include modality alignment. Cell-type prediction on chromatin accessibility data only has 0.285 macro F1, indicating a discrepancy between chromatin accessibility and gene expression data. We then added noise contrastive estimation to further boost unsupervised representation learning for chromatin accessibility. We observed a jump in improvement from the baseline model, and it suggests that learning structure of chromatin accessibility data, though without modality alignment, can also improve model. Next, we added adversarial domain loss into the training. We further observed a improvement from 0.440 to 0.480. Only after we included cylce-consistency loss jumped macro F1 to 0.525. Finally, we obtained the highest macro F1 when we further included ICT into the model. Based on results above, we concluded our final model as shown in Figure 1.

\begin{figure*}[t!]
    \includegraphics[width=14cm,height=18cm,page=4]{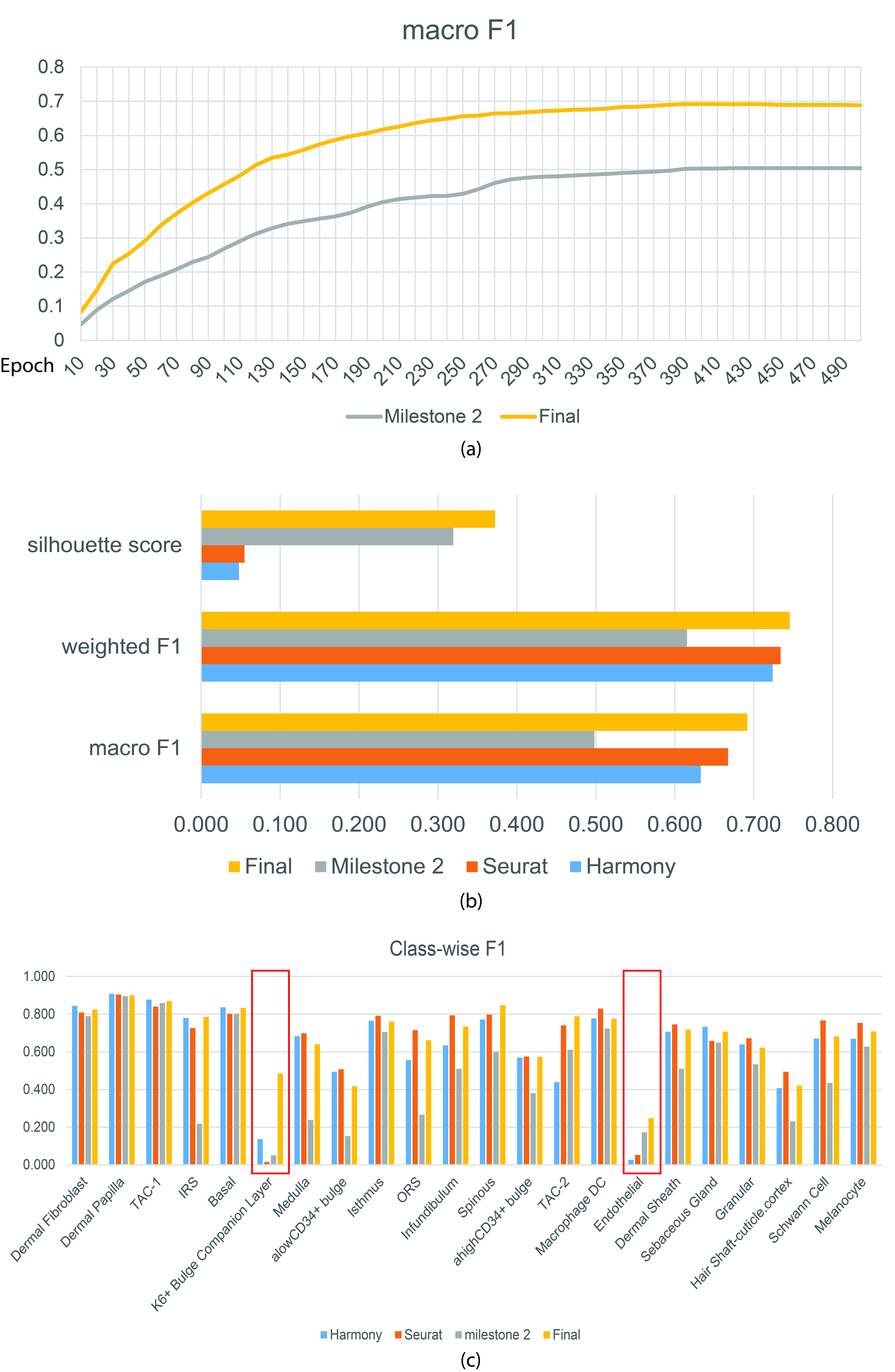}
    \centering
    \caption{(a) Learning curve. Macro F1 was calculated with unlabeled ATAC-seq data. Model prediction is against authors' cell-type annotation. Our final model was compared with model of milestone 2. Both models were trained for 500 epochs. (b) Evaluation of integration. Our final model was compared to milestone 2 model, Harmony, and Seurat. Harmony and Seurat are two SOTA methods for removing non-biological variations. In this study, Harmony and Seurat were used to remove modality difference in order to integrate chromatin accessibility and gene expression data. For all three metrics, higher value indicates better integration and higher accuracy of transfer learning. (c) Comparison at cell-type class level. Our final model outperforms all three methods in two particular cell types, in terms of more accurate cell-type prediction (marked in red boxes).}
\end{figure*}

\begin{figure*}[t!]
    \includegraphics[width=\textwidth,page=5]{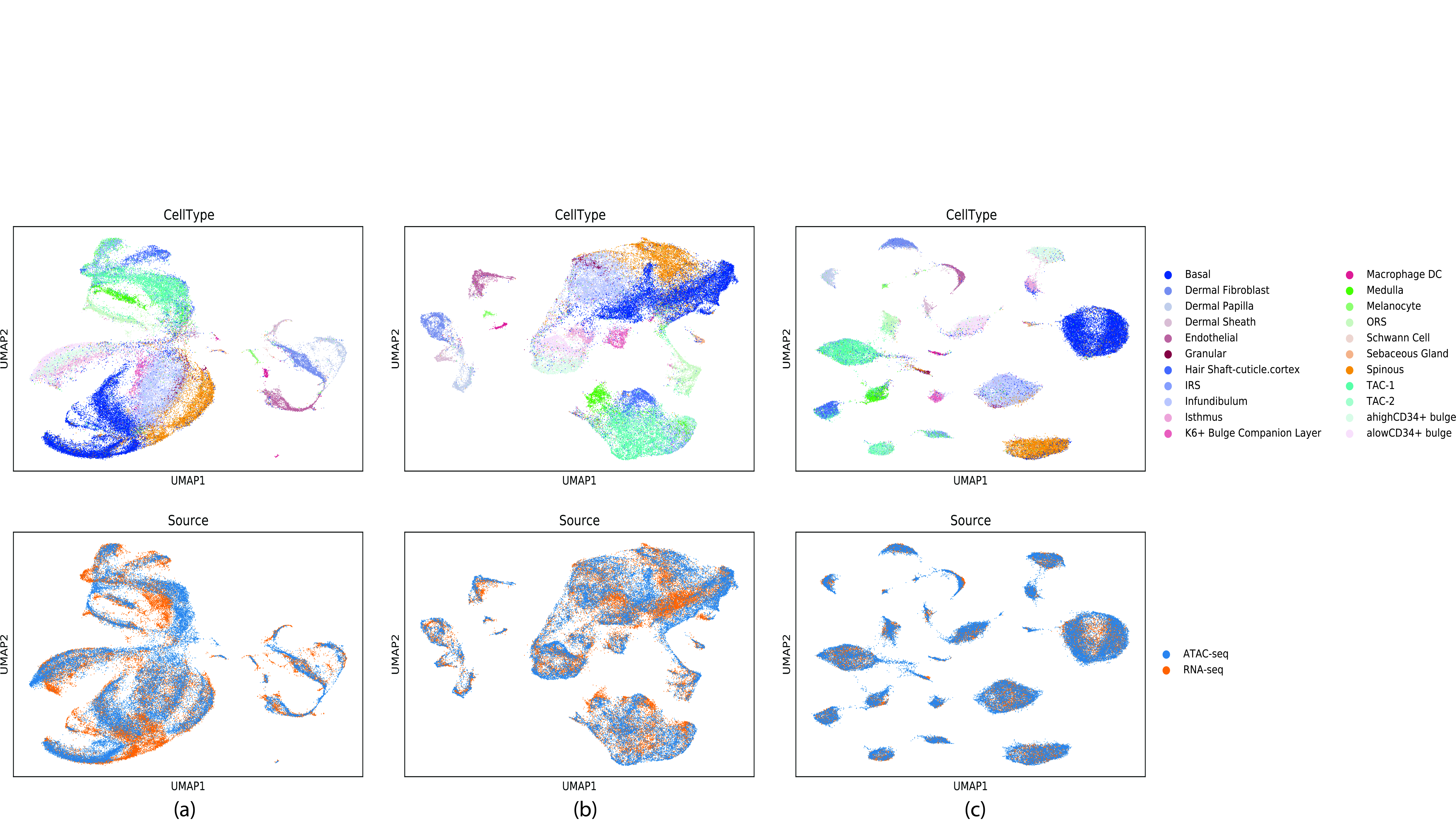}
    \centering
    \caption{(a) Data integration via Harmony; (b) via Seurat; (c) via our method. Cells are visualized through 2d UMAP and are colored by authors' annotation labels (in the upper panel) or according to its modality type (in the lower panel).}
\end{figure*}

\subsubsection{Improvement}
Our milestone 2 model is the same as our final model that uses gene expression data for supervised learning. However, milestone 2 model is trained with a simple adversarial domain adaption for modality alignment. We didn't use chromatin accessibility data for unsupervised learning in the milestone 2 model. To further boost discriminative representation learning, we alternatively used center loss to push cells in the latent space closer to class centroids\cite{wen2016discriminative}. Meanwhile, we applied center loss to cells in chromatin accessibility data that C has high confidence (> 0.8) in cell-type prediction. Doing this can assist modality alignment. Figure 2(a) shows that our final model substantially improve accuracy of integration from our milestone 2 model in human PBMC data, in terms of all three evaluation metrics. In mouse skin data, we trained both our milestone 2 model and final model for 500 epochs and monitored cell-type prediction of classifier on chromatin accessibility data. Figure 3(a) demonstrates that learning in our final model reaches to 0.7 macro F1 score and is also overwhelmingly boosted.

\subsubsection{Comparison with SOTA methods}
We next compared our methods with Harmony and Seurat, which are two SOTA methods for integration of single-cell sequencing data\cite{tran2020benchmark}. Harmony uses PCA-reduced data to iteratively align up the same cell clusters in different batches or modalities\cite{korsunsky2019fast}, while Seurat first identifies high-confident pairs and performs canonical correlation analysis to align up modalities\cite{stuart2019comprehensive}. Figure 3(b) shows that our method more correctly predicts cell-types for chromatin accessibility data than Harmony and Seurat in mouse skin data (macro and weighted F1 scores). Silhouette score indicates how well the integrate representation preserves data structure of mouse skin data. Our methods, including the milestone 2 model, overwhelmingly surpass both Harmony and Seurat. However, we have to acknowledge that our methods take advantage of supervised learning while both Harmony and Seurat are fully-unsupervised. Meanwhile, we examined label transferring from gene expression to chromatin accessibility at the class level. In Figure 3(c), we noticed that our methods greatly improved prediction accuracy on two particular cell-types, for which both Harmony and Seurat almost fails prediction. We also used UMAP to visualize the integrated representation. Integrated representation via Harmony, Seurat, and our final model were further reduced to 2d UMAP space for visualization. We colored cells from both chromatin accessibility and gene expression either according to authors' annotation label or modality source. Both Harmony and Seurat did not diminish the modality difference in Figure 4 (a) and (b), but our method greatly mixed cells from both modalities while preserving the distinct structure of cellular composition in Figure 4 (c). This also explains the higher silhouette score of our method.

\section{Summary}
In this study, we devised a novel adversarial approach for integration of unpaired single-cell chromatin accessibility and gene expression data. Our method is able to train an encoder to approximate joint representation for cells measured through single-cell ATAC-seq and RNA-seq, respectively. We systematically examined how different loss functions contribute to the integrated representation learning by using human PBMC data, and our final model substantially improved accuracy of integration from our milestone 2 method. Finally, we demonstrated that our adversarial approach outperformed two SOTA methods in mouse skin data, in terms of 3 different evaluation metrics.

\section{Future Work}
Results in our study indicate a promising use of adversarial learning in integration of multi-modal single-cell data, and we envision our method can be generalized for a broader use. However, a more extensive examination on single-cell data from a broader range of biological systems should be done in the near future. Meanwhile, our study mainly focused on integrating single-cell chromatin accessibility and gene expression data. We realize that generalizing our method for integration of other types of single-cell data will be extremely helpful for research community, and we aim to examine if our method is also applicable in these scenarios.

\nocite{*}
\bibliographystyle{acm}
\bibliography{main}

\begin{thebibliography}{10}

\bibitem{arisdakessian2019deepimpute}
{\sc Arisdakessian, C., Poirion, O., Yunits, B., Zhu, X., and Garmire, L.~X.}
\newblock Deepimpute: an accurate, fast, and scalable deep neural network
  method to impute single-cell rna-seq data.
\newblock {\em Genome biology 20}, 1 (2019), 1--14.

\bibitem{bahrami2020deep}
{\sc Bahrami, M., Maitra, M., Nagy, C., Turecki, G., Rabiee, H., and Li, Y.}
\newblock Deep feature extraction of single-cell transcriptomes by generative
  adversarial network.
\newblock {\em bioRxiv\/} (2020).

\bibitem{bernstein2020solo}
{\sc Bernstein, N.~J., Fong, N.~L., Lam, I., Roy, M.~A., Hendrickson, D.~G.,
  and Kelley, D.~R.}
\newblock Solo: doublet identification in single-cell rna-seq via
  semi-supervised deep learning.
\newblock {\em Cell Systems 11}, 1 (2020), 95--101.

\bibitem{cao2018joint}
{\sc Cao, J., Cusanovich, D.~A., Ramani, V., Aghamirzaie, D., Pliner, H.~A.,
  Hill, A.~J., Daza, R.~M., McFaline-Figueroa, J.~L., Packer, J.~S.,
  Christiansen, L., et~al.}
\newblock Joint profiling of chromatin accessibility and gene expression in
  thousands of single cells.
\newblock {\em Science 361}, 6409 (2018), 1380--1385.

\bibitem{chen2020simple}
{\sc Chen, T., Kornblith, S., Norouzi, M., and Hinton, G.}
\newblock A simple framework for contrastive learning of visual
  representations.
\newblock In {\em International conference on machine learning\/} (2020), PMLR,
  pp.~1597--1607.

\bibitem{dincer2020adversarial}
{\sc Dincer, A.~B., Janizek, J.~D., and Lee, S.-I.}
\newblock Adversarial deconfounding autoencoder for learning robust gene
  expression embeddings.
\newblock {\em Bioinformatics 36}, Supplement\_2 (2020), i573--i582.

\bibitem{eraslan2019single}
{\sc Eraslan, G., Simon, L.~M., Mircea, M., Mueller, N.~S., and Theis, F.~J.}
\newblock Single-cell rna-seq denoising using a deep count autoencoder.
\newblock {\em Nature communications 10}, 1 (2019), 1--14.

\bibitem{ganin2015unsupervised}
{\sc Ganin, Y., and Lempitsky, V.}
\newblock Unsupervised domain adaptation by backpropagation.
\newblock In {\em International conference on machine learning\/} (2015), PMLR,
  pp.~1180--1189.

\bibitem{gawad2016single}
{\sc Gawad, C., Koh, W., and Quake, S.~R.}
\newblock Single-cell genome sequencing: current state of the science.
\newblock {\em Nature Reviews Genetics 17}, 3 (2016), 175.

\bibitem{kimmel2020scnym}
{\sc Kimmel, J.~C., and Kelley, D.~R.}
\newblock scnym: Semi-supervised adversarial neural networks for single cell
  classification.
\newblock {\em bioRxiv\/} (2020).

\bibitem{korsunsky2019fast}
{\sc Korsunsky, I., Millard, N., Fan, J., Slowikowski, K., Zhang, F., Wei, K.,
  Baglaenko, Y., Brenner, M., Loh, P.-r., and Raychaudhuri, S.}
\newblock Fast, sensitive and accurate integration of single-cell data with
  harmony.
\newblock {\em Nature methods 16}, 12 (2019), 1289--1296.

\bibitem{lahnemann2020eleven}
{\sc L{\"a}hnemann, D., K{\"o}ster, J., Szczurek, E., McCarthy, D.~J., Hicks,
  S.~C., Robinson, M.~D., Vallejos, C.~A., Campbell, K.~R., Beerenwinkel, N.,
  Mahfouz, A., et~al.}
\newblock Eleven grand challenges in single-cell data science.
\newblock {\em Genome biology 21}, 1 (2020), 1--35.

\bibitem{lee2019simultaneous}
{\sc Lee, D.-S., Luo, C., Zhou, J., Chandran, S., Rivkin, A., Bartlett, A.,
  Nery, J.~R., Fitzpatrick, C., O’Connor, C., Dixon, J.~R., et~al.}
\newblock Simultaneous profiling of 3d genome structure and dna methylation in
  single human cells.
\newblock {\em Nature methods 16}, 10 (2019), 999--1006.

\bibitem{li2019joint}
{\sc Li, G., Liu, Y., Zhang, Y., Kubo, N., Yu, M., Fang, R., Kellis, M., and
  Ren, B.}
\newblock Joint profiling of dna methylation and chromatin architecture in
  single cells.
\newblock {\em Nature methods 16}, 10 (2019), 991--993.

\bibitem{lopez2018deep}
{\sc Lopez, R., Regier, J., Cole, M.~B., Jordan, M.~I., and Yosef, N.}
\newblock Deep generative modeling for single-cell transcriptomics.
\newblock {\em Nature methods 15}, 12 (2018), 1053--1058.

\bibitem{lotfollahi2018generative}
{\sc Lotfollahi, M., Wolf, F.~A., and Theis, F.~J.}
\newblock Generative modeling and latent space arithmetics predict single-cell
  perturbation response across cell types, studies and species.
\newblock {\em bioRxiv\/} (2018), 478503.

\bibitem{ma2020actinn}
{\sc Ma, F., and Pellegrini, M.}
\newblock Actinn: automated identification of cell types in single cell rna
  sequencing.
\newblock {\em Bioinformatics 36}, 2 (2020), 533--538.

\bibitem{ma2020chromatin}
{\sc Ma, S., Zhang, B., LaFave, L.~M., Earl, A.~S., Chiang, Z., Hu, Y., Ding,
  J., Brack, A., Kartha, V.~K., Tay, T., et~al.}
\newblock Chromatin potential identified by shared single-cell profiling of rna
  and chromatin.
\newblock {\em Cell 183}, 4 (2020), 1103--1116.

\bibitem{stuart2019comprehensive}
{\sc Stuart, T., Butler, A., Hoffman, P., Hafemeister, C., Papalexi, E.,
  Mauck~III, W.~M., Hao, Y., Stoeckius, M., Smibert, P., and Satija, R.}
\newblock Comprehensive integration of single-cell data.
\newblock {\em Cell 177}, 7 (2019), 1888--1902.

\bibitem{stuart2019integrative}
{\sc Stuart, T., and Satija, R.}
\newblock Integrative single-cell analysis.
\newblock {\em Nature Reviews Genetics 20}, 5 (2019), 257--272.

\bibitem{tran2020benchmark}
{\sc Tran, H. T.~N., Ang, K.~S., Chevrier, M., Zhang, X., Lee, N. Y.~S., Goh,
  M., and Chen, J.}
\newblock A benchmark of batch-effect correction methods for single-cell rna
  sequencing data.
\newblock {\em Genome biology 21}, 1 (2020), 1--32.

\bibitem{tzeng2017adversarial}
{\sc Tzeng, E., Hoffman, J., Saenko, K., and Darrell, T.}
\newblock Adversarial discriminative domain adaptation.
\newblock In {\em Proceedings of the IEEE conference on computer vision and
  pattern recognition\/} (2017), pp.~7167--7176.

\bibitem{verma2019interpolation}
{\sc Verma, V., Kawaguchi, K., Lamb, A., Kannala, J., Bengio, Y., and
  Lopez-Paz, D.}
\newblock Interpolation consistency training for semi-supervised learning.
\newblock {\em arXiv preprint arXiv:1903.03825\/} (2019).

\bibitem{wang2019bermuda}
{\sc Wang, T., Johnson, T.~S., Shao, W., Lu, Z., Helm, B.~R., Zhang, J., and
  Huang, K.}
\newblock Bermuda: a novel deep transfer learning method for single-cell rna
  sequencing batch correction reveals hidden high-resolution cellular subtypes.
\newblock {\em Genome biology 20}, 1 (2019), 1--15.

\bibitem{wen2016discriminative}
{\sc Wen, Y., Zhang, K., Li, Z., and Qiao, Y.}
\newblock A discriminative feature learning approach for deep face recognition.
\newblock In {\em European conference on computer vision\/} (2016), Springer,
  pp.~499--515.

\bibitem{zhu2017unpaired}
{\sc Zhu, J.-Y., Park, T., Isola, P., and Efros, A.~A.}
\newblock Unpaired image-to-image translation using cycle-consistent
  adversarial networks.
\newblock In {\em Proceedings of the IEEE international conference on computer
  vision\/} (2017), pp.~2223--2232.

\end{thebibliography}


\end{document}